\documentclass[manuscript]{aastex}
\usepackage[latin1]{inputenc}
\usepackage{spr-astr-addons}
\usepackage{url}

\usepackage{graphicx}	
\usepackage{amsmath}	
\usepackage{amssymb}	
\usepackage{lineno}
\usepackage{multirow}
\usepackage{diagbox}
\usepackage{subfigure}
\usepackage[colorlinks = true, linkcolor = blue, anchorcolor = red, citecolor = black, filecolor = red, pagecolor = red, urlcolor = blue]{hyperref}

\RequirePackage{color}

\newcommand{\emaila}{authors@email.com}

\begin{document}
\begin{sloppypar}

\title{A Hybrid Ensemble method for Pulsar Candidate Classification}
\slugcomment{Not to appear in Nonlearned J., 45.}
\shorttitle{Pulsar Candidate Classification}
\shortauthors{Wang et al.}

\author{Y. Wang\altaffilmark{1,2}}
\and
\author{Z. Pan\altaffilmark{3,4,5}}
\and
\author{J. Zheng\altaffilmark{1,2}}
\and
\author{L. Qian\altaffilmark{3,4,5}}
\and
\author{M. Li\altaffilmark{1,2}}
\affil{\email{\emaila}{limingtao@nssc.ac.cn}}

\altaffiltext{1}{National Space Science Center, Chinese Academy of Sciences, Beijing 100190, China}
\altaffiltext{2}{University of Chinese Academy of Sciences, Beijing 10004, China}
\altaffiltext{3}{National Astronomical Observatories, Chinese Academy of Sciences, Beijing 10012, China}
\altaffiltext{4}{Center for Astronomical Mega-Science, Chinese Academy of Sciences, Beijing, 100012, China}
\altaffiltext{5}{CAS Key Laboratory of FAST, NAOC, Chinese Academy of Sciences, Beijing, 100012, China}

\begin{abstract}
  In this paper, three ensemble methods: Random Forest, XGBoost, and a Hybrid Ensemble method were implemented to classify imbalanced pulsar candidates.
  To assist these methods, tree models were used to select features among 30 features of pulsar candidates from references.
  The skewness of the integrated pulse profile, chi-squared value for sine-squared fit to amended profile and best S/N value play important roles in Random Forest,
  while the skewness of the integrated pulse profile is one of the most significant features in XGBoost.
  More than 20 features were selected by their relative scores and then applied in three ensemble methods.
  In the Hybrid Ensemble method, we combined Random Forest and XGBoost with EasyEnsemble.
  By changing thresholds, we tried to make a trade-off between Recall and Precision to make them approximately equal and as high as possible.
  Experiments on HTRU 1 and HTRU 2 datasets show that the Hybrid Ensemble method achieves higher Recall than the other two algorithms.
  In HTRU 1 dataset, Recall, Precision, and F-Score of the Hybrid Ensemble method are $0.967$, $0.971$, and $0.969$, respectively.
  In HTRU 2 dataset, the three values of that are $0.920$, $0.917$, and $0.918$, respectively.

\end{abstract}


\section{Introduction}

  Most radio pulsars were discovered through dedicated surveys,
  including Parkes Multi-beam Pulsar Survey (PMPS) \citep{Manchester2001},
  High Time Resolution Universe (HTRU) Parkes survey  \citep{Burke-Spolaor2011},
  Pulsar Arecibo L-band Feed Array survey (PALFA) \citep{Deneva2009},
  the Green Bank Northern Celestial Cap pulsar survey (GBNCC)\citep{Stovall2014},
  and the Commensal Radio Astronomy FAST (the Five-hundred-meter Aperture Spherical Telescope, \citealt{Nan2011}) Survey (CRAFTS)\footnote{\url{http://crafts.bao.ac.cn/pulsar/}}, etc..
  These surveys searched for  periodic signals
  and the search results were reduced into diagnostic values and/or graphical representations for researchers selecting pulsar candidates.
  With search techniques and telescope sensitivities being improved, more and more candidates are obtained.
  Most candidates arise from noise or radio frequency interference (RFI)\citep{Lyon2016why}.
  As a particular example, a pulsar survey undertaken with the Square Kilometre Array (SKA) \citep{Smits2009} is expected to detect 20000 pulsars with more than 200 million candidates in a conservative assumption \citep{Lyon2013}.
  It is infeasible to classify candidates manually.
  Many successful techniques have been developed to find candidates with pulsar-like features among the search results.

  One of these methods uses image processing tools.
  Designed graphical selection tools, for example, Reaper \citep{Faulkner2004}, aid researchers selecting candidates.
  The Reaper presents each candidate as a single point, and presents thousands of candidates onto plots.
  Since the distribution of pulsar candidates differ from that of others (such as RFIs) on those plots,
  part of pulsar candidates can be easily identified.
  Following the success of Reaper, JReaper was developed by \citet{Keith2009}.
  It is a heuristic scoring method based on specified weights, which makes candidates ranking more sufficient.
  \citet{Lee2013} developed a scoring algorithm named PEACE,
  which uses six features to determine candidates scores with entirely predetermined simple functions and linear combinations.

  Machine learning classifiers were also used.
  \citet{Eatough2010} used machine learning approaches for pulsar search.
  In this work, they selected 12 features as input vectors and used artificial neural networks (ANN) to output candidate scores.
  It was applied to reanalysis PMPS data and discovered one new pulsar.
  Based on \citet{Eatough2010}, \citet{Bates2012} added 10 new features and
  used the single-hidden-layer ANN to classify both millisecond pulsars and normal pulsars in HTRU data.
  Eighty-five percent of pulsars were detected while 99\% candidates were rejected.
  Straightforward Pulsar Identification using Neural Networks (SPINN) system,
  which trains an artificial neural network only with six features, was used in candidates selection by \citet{Morello2014}.
  They found four new pulsars in a re-processing of the HTRU data.
  Based on feature of  \citet{Morello2014},
  \citet{Bethapudi2018} applied four different algorithms with Synthetic Minority Over-sampling Technique (SMOTE) and analysis of feature importances to make progress for this problem.
 
  Differently, \citet{Zhu2014} developed Pulsar Image-based Classification System(PICS) system by using a group of supervised machine learning approaches,
  including ANN, convolutional neural network (CNN), support vector machine (SVM), and logistic regression.
  As an improvement, PICS applies image pattern recognition.
  The inputs are diagnostic plots of candidates rather than extracted features in  ANN-based methods.
  Considering the need for online processing, \citet{Lyon2016} presented a new method using a purpose-built tree-based machine learning classifier,
  Gaussian Hellinger very fast decision tree(GH-VFDT), and chose eight new features without intrinsic biases.
  The features of \citet{Lyon2016} were used by \citet{Mohamed2017}.
  The fuzzy-K nearest Neighbors (knn) classifier was used to select pulsar candidates in HTRU data.
   \citet{Tan2018} introduced 12 new features and a third class characterizing RFI instance.
  Five decision trees were used to develop an ensemble classifier.
  It is being implemented into the LOFAR Tied-Array All-sky Survey(LOTAAS) search pipeline.

  Those machine learning methods improved the classification performance, but there is still something for further improvement.
  Some features may be sub-optimal \citep{Lyon2016,Tan2018}.
  Besides, we found that those existing methods still returned too many false positives. For example, in HTRU2 dataset, fuzzy-knn has improved recall to be 0.942, but the precision was 0.808\citep{Mohamed2017}.
  For more complex real-world data, precision may be lower.
  We hope to adopt optimized features and improved classification methods, to identify more pulsar candidates and reduce the false positives at the same time.
  In this paper, we introduce our pulsar candidates features selections, the three used ensemble algorithms,
  which are Random Forest \citep{Breiman2001}, XGBoost \citep{Chen2016}, and a Hybrid Ensemble method,
  and tested results on HTRU 1 and HTRU 2 datasets.
  The following parts are organized as follows:
  In Section 2, we describe the pulsar candidates and basic features.
  In Section 3, we mainly introduce the three fundamental models and performance metrics.
  In Section 4, we describe the approaches for feature selection and analyse the relative importance of features.
  The experimental results on two pulsar datasets and discussion are in Section 5 and 6.
  Conclusions are given in section 7.

\section{Pulsar candidates and features}

\begin{figure}[t]
      \includegraphics[width=\columnwidth]{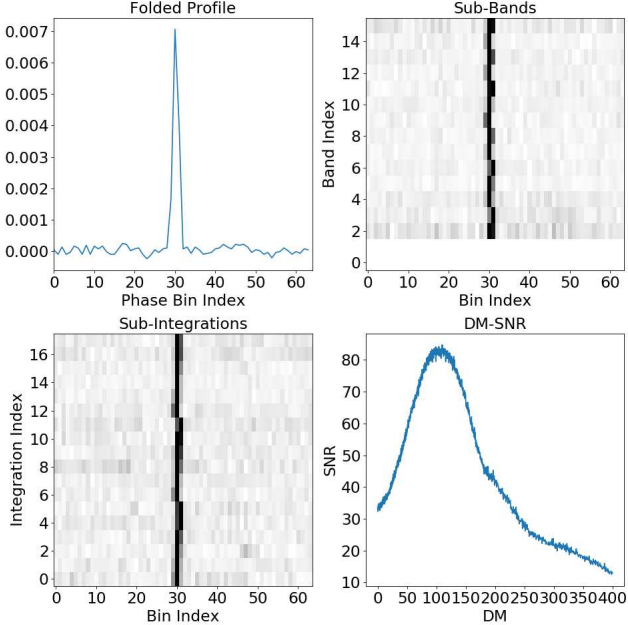}
      \caption{An example (Pulsar\_0003) from HTRU 1 dataset.
      These four subplots are: folded profile plot(upper left), sub-integrations plot(lower left), sub-bands plot(upper right), and DM-SNR curve(lower right).}
      \label{fig:Figure 1}
\end{figure}

  The characteristics of a signal detection can be visualized by many different diagnostic plots.
  The typical subplots can be folded profile plot, sub-integrations plot, sub-bands plot, and DM-SNR curve.
  Most features are extracted from them.
  Figure~\ref{fig:Figure 1} is an example of a candidate from HTRU 1 dataset.
  The upper left is the folded profile plot, it is an intensity vs phase plot, which is obtained by summing all frequency channels and time intervals.
  The lower left is the sub-integrations plot, it is obtained by summing data of different frequency channels while summing data over all time intervals make the upper right one, the sub-bands plot.
  In the DM-SNR curve at lower right, S/N ratio as a function of DM is recorded.

  A candidate, which is believed to be a real pulsar,
  should have at least one or more features meet the requirements as
  1) one or several peaks in folded profile plot,
  2,3) several persistent vertical lines correspond to the peaks in sub-integrations plot and/or sub-bands plot,
  and 4) a well-defined maximum value on the DM curve.
  Our selected features are used to characterize them.

\subsection{Datasets}
\label{sec:Datasets} 

  The datasets we used here are HTRU 1 and HTRU 2, which were public and can be downloaded from related websites.
  The HTRU 1 dataset\footnote{\url{http://astronomy.swin.edu.au/~vmorello/}} was produced by \citet{Morello2014}.
  It is the first public dataset with labeled candidates, which consists of 1,196 pulsar  and 89,995 non-pulsar candidates.
  This dataset is a part of outputs of new processing of HTRU medlat data and contains pulsars with varied spin periods, duty cycles, and signal to noise ratios \citep{Morello2014}.
  \citet{Lyon2016} made the HTRU 2 dataset \footnote{\url{https://figshare.com/articles/HTRU2/3080389/1}} available.
  It consists of 1,639 pulsar candidates and 16,259 non-pulsar candidates,
  which are obtained from an analysis of HTRU Medium Latitude data by \citet{Thornton2013}.

  It should be noted that the datasets are obviously imbalanced, which may affect the performance of classifiers.

\subsection{Feature Selection}
\label{sec:Feature Selection} 
  Researchers have proposed many features to characterize pulsar candidates,
  and some of them have been proven to be effective.
  For HTRU 1 and 2 data, the 22 features in \citet{Bates2012} and the eight features in \citet{Lyon2016} are listed in Table~\ref{tab:Table 1} ,which are more commonly used.
  It is worthy to point out that although the 8 features in \citet{Lyon2016} have no intrinsic biases and perform well \citep{Lyon2016, Mohamed2017},
  they are only obtained from the folded profile and DM-SNR curve, which may miss some information.
  In addition, some of those features have been turned out to be sub-optimal or have no predictive value in some cases\citep{Tan2018, Ford2017}.
  In order to construct a more suitable feature sets, we combined 8 features in \citet{Lyon2016} and 22 features in \citet{Bates2012}, then selected important features.
  All these 30 features were extracted with the help of a publicly available python tool, PULSAR FEATURE LAB \footnote{\url{https://figshare.com/articles/Pulsar_Feature_Lab/1536472}}, produced by \citet{Lyon2016}.


\begin{table*}[t]
	\centering
	\caption{The features that we chose to test are from two parts: number 1 to 8 are from \citet{Lyon2016} while number 9 to 30 are from \citet{Bates2012}.}
	\label{tab:Table 1}
	\begin{tabular}{lccr} 
		\hline
		Number &	Describe of Features\\
        \hline
               &    From \citet{Lyon2016} \\
        \hline
        1	& Mean of the integrated (folded) pulse profile.\\
        2	& Standard deviation of the integrated (folded) pulse profile.\\
        3	& Skewness of the integrated (folded) pulse profile.\\
        4	& Excess kurtosis of the integrated (folded) pulse profile.\\
        5	& Mean of the DM-SNR curve.\\
        6	& Standard deviation of the DM-SNR curve.\\
        7	& Skewness of the DM-SNR curve.\\
        8	& Excess kurtosis of the DM-SNR curve.\\
        \hline
            & From \citet{Bates2012} \\
        \hline
        9	& Chi-Squared value for sine fit to raw profile.\\
        10	& Chi-Squared value for sine-squared fit to amended profile.\\
        11	& Number of peaks the program identifies in the pulse.\\
        12	& Area under the pulse profile after subtracting mean.\\
        13	& Distance between expectation values of Gaussian and fixed Gaussian fits to profile histogram.\\
        14	& Ratio of the maximum values of Gaussian and fixed Gaussian fits to profile histogram.\\
        15	& Distance between expectation values of derivative histogram and profile histogram.\\
        16	& Full-width-half-maximum (FWHM) of Gaussian fit to pulse profile.\\
        17	& Chi squared value from Gaussian fit to pulse profile.\\
        18	& Smallest FWHM of double-Gaussian fit to pulse profile.\\
        19	& Chi squared value from double Gaussian fit to pulse profile.\\
        20	& Best period.\\
        21	& Best S/N value.\\
        22	& Best DM value.\\
        23	& Best pulse width.\\
        24	& $SNR /\sqrt{(P-W)/W}$ .\\
        25	& $SNR_{fit}  /\sqrt{(P-W)/W}$.\\
        26	& $mod(DM_{fit}-DM_{best})$.\\
        27	& Chi squared value from DM curve fit.\\
        28	& RMS of peak positions in all sub-bands.\\
        29	& Average correlation coefficient for each pair of sub-bands.\\
        30	& Sum of correlation coefficients.\\
		\hline
	\end{tabular}
\end{table*}


\section{Ensemble learning}

  Ensemble learning is one class of machine learning methods that trains multiple learners to solve the same problem and combines them to obtain a new better learner.
  These multiple learners are usually called base learners.
  When applying ensemble methods,
  base learners can be different models (such as decision trees, neural networks, support vector machines, or other learning algorithms) or same models but trained with different data/parameters.
  Rather than selecting the best single learner, final results of the ensemble can be determined by using averaging, voting, or stacking method.
  These processes increase the differentiation of ensemble models and reduce the risk of overfitting.

  Compared to the single model leaner with only one hypothesis over the data, ensemble learning can maintain multiple hypotheses.
  The generalization ability of an ensemble is often much stronger than that of base learners \citep{Zhou2012}.
  Ensemble learning is also one of the approaches for handling class imbalance \citep{Galar2012,Blaszczynski2015,Krawczyk2016}.
  \cite{Dietterich2000} gave three reasons to explain why ensemble methods usually perform better.
  Firstly, Ensemble methods with multiple hypotheses can reduce the risk of choosing a wrong hypothesis or being underfitting, especially when there are many learners perform equally well on the limited training dataset.
  Secondly,  many algorithms may fall in local optima.
  In cases of enough training data, it may still be tough to find the best hypothesis.
  Ensemble methods may provide a better approximation to the true unknown hypotheses and reduce the risk of local optima.
  Thirdly, in some cases, the hypothesis space being searched might not contain the true target function. If using a single model leaner, it will be useless. By combining some hypotheses, it may be possible to expand the space and make ensemble methods form a better approximation.

  Two families of ensemble methods are usually distinguished: bagging methods and boosting methods.
  In bagging methods, such as Bagging and Random Forest,
  the driving principle is to build several parallel base learners and then to form a final result by averaging those predictions. It reduces the variance.
  By contrast, in boosting methods, such as AdaBoost, Gradient Tree Boosting, and Extreme Gradient Boosting (XGBoost),
  the purpose is to combine several weak base learners to produce a powerful learner. Base learners are built sequentially, and final results are summed together trying to reduce the bias.

  In these methods, Random Forest and XGBoost are used widely.
  They are simple, efficient, easy to implement and usually have excellent generalization performance, which are suitable for dealing with a large number of pulsar candidates.
  Both Random Forest and XGBoost can automatically provide estimates of feature importance from a trained predictive model, which can help us to select feature set \citep{Genuer2010,Hira2015,Xiao2017,Mangal2016}.
  So we chose Random Forest and XGBoost to classify candidates.

  Before using them, we should solve the problem of imbalanced pulsar dataset.
  These methods for learning from imbalanced data may be distinguished into three main classes: data level methods, algorithm level methods, and hybrid methods.
  It is proved that hybridization of bagging and boosting with sampling or cost-sensitive methods \citep{Elkan2001,Zhou2010} are highly competitive and robust to difficult data \citep{Krawczyk2016}.
  So considering these, we chose and tested three methods:

  (1)	Random Forest with adopting a cost-sensitive approach to address imbalance class problem ;

  (2)	XGBoost with adopting a cost-sensitive approach to address imbalance class problem;

  (3)	A Hybrid Ensemble method combining Random Forest and XGBoost with EasyEnsemble \citep{Liu2009}.


\subsection{Decision tree}

  Decision tree is a kind of common machine learning algorithm, whose purpose is to create a model that predicts the value of a target variable by learning simple decision rules inferred from the data features.
  The model is a tree-like structure.
  In general, each tree model contains a root node, multiple internal nodes, and multiple leaf nodes.
  The root node contains all samples,
  then it divides data into child internal nodes according to the optimal attribute selected by given criterion (such as information entropy, grain ratio, Gini index).
  Each internal node will divide continually, until most data are classified correctly.
  Ideally, samples contained in each leaf node can be classified into the same category as far as possible.
  Therefore, the path from the root to each leaf corresponds to a decision test sequence, which is a divide-and-conquer strategy, and leaf nodes corresponding to decision results.

  The decision tree algorithm is the basis of Random Forest, XGBoost, and other tree models.

\subsection{Random Forest}
\label{sec:Random Forest} 

\begin{figure}[t]
      \includegraphics[width=\columnwidth]{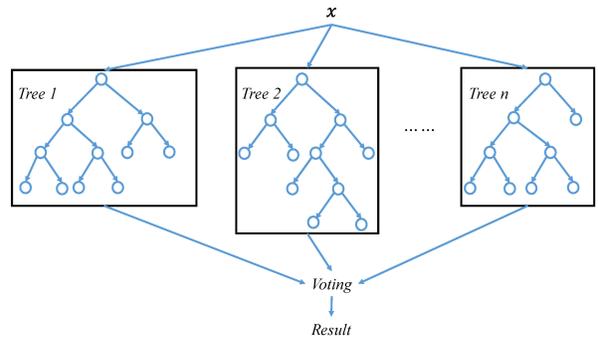}
      \caption{A general architecture of Random Forest.}
      \label{fig:Figure 2}
\end{figure}

  Random Forest\footnote{\url{http://scikit-learn.org/stable/modules/ensemble.html\#forest}} is a variant of bagging method.
  Its base learners are classification regression trees. Figure~\ref{fig:Figure 2} shows a general architecture of Random Forest.
  The steps to generate a Random Forest are as follows :

  1)  From the original training dataset,
      retrieving $k$ subsets randomly with the bootstrap method (left samples are not drawn but to form out-of-bag(OOB) data).
      Here, $k$ is an integer, and these $k$  subsets are to build $k$ classification regression trees.

  2) Growing a tree $Ta$ based on the subset $a$ by recursively repeating the following for each non-leaf nodes until the tree grows to the maximum size:
     A random subset of features is selected,
     and the optimal feature in this subset is picked by computing information amount. Then the node is split into two child nodes.

  3) Repeating the second step for $k$ time, then $k$ trees combine as the random forest, and the final results are determined by multiple tree classifiers using majority voting method.

In binary classification problem, majority voting method can be expressed as:
\begin{equation}
    \hat{y}=\begin{cases}
    1,& {\sum_{i=0}^{k}y_i>=\theta}, \\
    0,& {\sum_{i=0}^{k}y_i<\theta}
    \end{cases}
\end{equation}
where $\widehat{y}$ is the final result, $y_i$ is i-th tree's result and $\theta$ is a voting threshold.

  In Random Forest, each tree is built from a random subset of the training set using bootstrap sampling like other normal bagging methods.
  In addition, there also introduces the randomness of features.
  When splitting a node during the construction of the tree,
  normal bagging method would choose the best split among all features.
  In Random Forest, it picks the best split among a random subset of the features.
  As a result, these randomnesses increase the diversities of base trees, decrease variance, and yield an overall better model.


\subsection{XGBoost}
  XGBoost\footnote{The Python package of  XGBoost is in: \url{https://github.com/dmlc/xgboost}} is optimized distributed gradient boosting system.
  It is used widely by data scientists to achieve state-of-the-art results on many machine learning challenges \citep{Chen2016}.

\begin{figure}[t]
\centering
\begin{center}
      \includegraphics[width=\columnwidth]{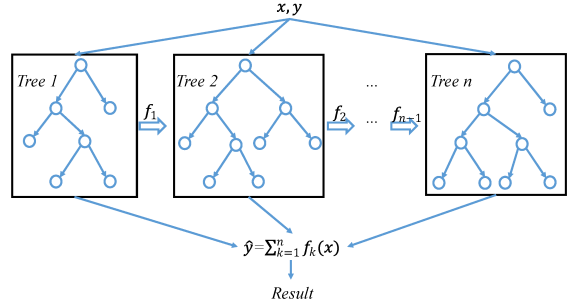}
      \caption{A general architecture of XGBoost.}
      \label{fig:Figure 3}
\end{center}
\end{figure}

  XGBoost is an iterative decision tree algorithm with multiple decision trees.
  Every tree is learning from the residuals of all previous trees.
  Rather than adopting most voting output results in Random Forest,
  the predicted output of XGBoost is the sum of all the results (shown in Figure~\ref{fig:Figure 3} ):

\begin{equation}
   \centering
    \hat{y_i}={\sum_{k=1}^{n}f_k(x_i)},\quad f_k \in F,
\end{equation}
  where $F$ means the space of regression trees, $f_k$ corresponds to a tree,
  so $f_k(x_i)$ is the result of tree $k$, and $\hat{y _i}$ is the predicted value of i-th instance $x_i$.

The objective of XGBoost is:
\begin{equation}
    Obj(\theta)=L(\theta)+\Omega(\theta),
\end{equation}
where $L(\theta)=\sum_{i=1}^{n}l(y_i,\hat{y_i})$  is loss function, $\hat{y_i}$ is the prediction and $y_i$ is the target, and $ \Omega(\theta)=\sum_{k=1}^{K}\Omega(f_k)$ penalizes the complexity of the model.

Then the model is trained in an additive manner. Letting ${\hat{y}_i}^{(t)}$ be the prediction of i-th instance at t-th iteration, ${\hat{y}_i}^{(t)}$ can be expressed as:
\begin{equation}
    {\hat{y}_i}^{(t)}={\hat{y}_i}^{(t-1)}+f_t (x_i).
\end{equation}
In this situation, it minimizes the following objective:
\begin{equation}
    Obj^{(t)}={\sum_{i=1}^{n}l(y_i,\hat{y}_i^{(t-1)}+f_t(x_i))+\Omega(f_t)}.
\end{equation}
In order to optimize the objective quickly, second-order approximation is used
\begin{equation}
    Obj^{(t)}={\sum_{i=1}^{n}(l(y_i,\hat{y}_i^{(t-1)})+g_if_t(x_i)+\frac{1}{2}h_if_t^2(x_i))+\Omega(f_t)},
\end{equation}
where $g_i$ and $h_i$ are first and second order gradient statistics on the loss function, respectively.

There are some skills to improve the performance, such as regularized boosting and column subsampling \citep[see more details in][]{Chen2016}. XGBoost is efficient, flexible, and has good generalization performance.

\subsection{Cost-sensitive methods}

   Cost-sensitive methods are algorithm-level methods. They concentrate on modifying existing learners by incorporating varying penalty for each of the considered groups to alleviate their bias towards majority groups\citep{Krawczyk2016}. The goal of this type of learning is to minimize the total cost.
   One process of these methods is to assign the training examples of different classes with different weights\citep{Zhou2010}.
   If $C(i, j)$ is  used to represent the misclassification cost of classifying an instance from its actual class $j$ into the predicted class $i$,  the cost matrix with two classes is shown in Table 2. The expected cost $R(i|x)$ of classifying an instance x into class $i$ (by a classifier) can be expressed as:
   \begin{equation}
    R(i|x)=\sum_{j}P(j|x)C(j,i)
  \end{equation}
  where $P(j|x)$ is the probability estimation of classifying an instance into class $j$.

  \begin{table}[t]
	\centering
	\caption{Cost matrix of binary classification.}
	\label{tab:Table 2}
	\begin{tabular}{ccc} 
        \hline
                   &    Predicted      &  Predicted \\
                   &    Positive       &   Negative\\	            		
        \hline
        Actual Positive	    &  C(1,1)           &  C(0,1)  \\
        Actual Negative    &   C(1,0)          &   C(0,0)  \\

        \hline
	\end{tabular}
\end{table}

  In our Random Forest and XGBoost model, to achieve this goal, we adjusted weighting parameters in algorithms for assigning different weights to pulsar candidates and non-pulsar candidates. We set $C(1,1)=C(0,0)=0$,  $C(1,0) = 1$, $C(0,1)=\frac{n}{m}$, where $n$ means the number of non-pulsar data and $m$ means the number of pulsar data.

\subsection{EasyEnsemble}
\begin{figure}[t]
      \includegraphics[width=\columnwidth]{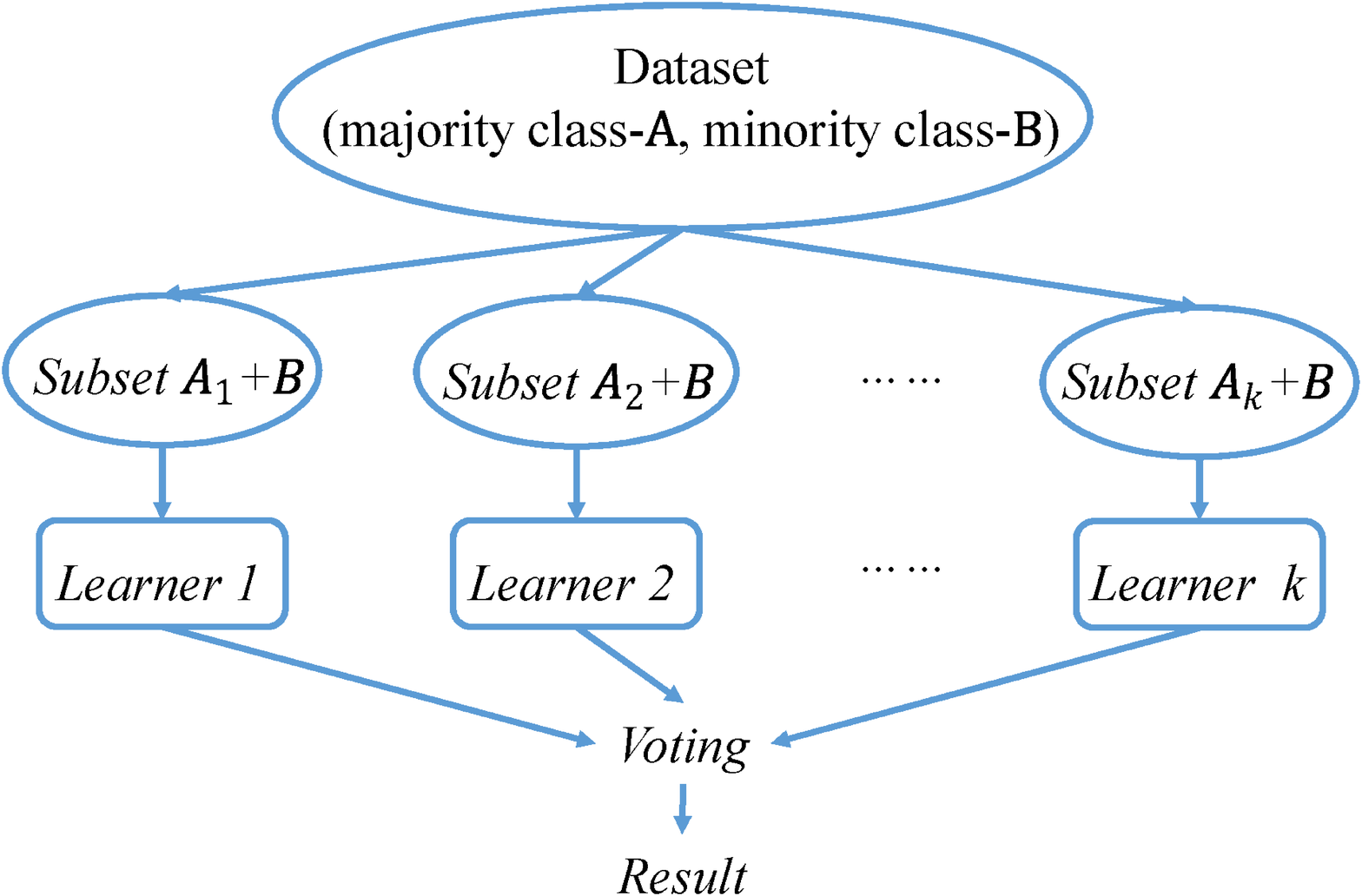}
      \caption{A general architecture of EasyEnsemble method. In our Hybrid Ensemble method, the base learner is Random Forest or XGBoost.}
      \label{fig:Figure 4}
\end{figure}

  In order to address the sample imbalance and further improve the stability of the model,
  we choose EasyEnsemble. EasyEnsemble is a Hybrid Ensemble method for class imbalance problem described by \citet{Liu2009}.
  It is an ensemble of ensembles.
  In EasyEnsemble, it would randomly extract $k$ subsets from the majority class (using the bootstrap method). In each subset, the number of data is equal to the number of data of minority class.
  Then it combines each subset with data of minority class together as balanced training data and trains $k$ learners respectively.
  Finally, these learners are combined to form an integrated learning system (see in Figure~\ref{fig:Figure 4}).

  It re-balances the class distribution by under-sampling for each base learner without losing any important information from the overall point of view, which makes EasyEnsemble stable.

  In normal EasyEnsemble, the base learner is AdaBoost classifiers. In this paper, in order to increase the diversity and improve the performance in identifying pulsar candidates, we choice Random Forest and XGBoost as base learners.

\subsection{Choice of performance metrics}

  Table 3 is the binary classification's confusion matrix.
  For binary classification problems, each predicted label is either correct or incorrect. True Positive (TP) represent those positive candidates correctly labeled. True Negative (TN) are those negative candidates correctly labeled. False Negative(FN) mean those positive candidates incorrectly labeled. False Positive (FP) are those negative candidates incorrectly labeled.

\begin{table}[t]
	\centering
	\caption{Confusion matrix of binary classification.}
	\label{tab:Table 3}
	\begin{tabular}{ccc} 
        \hline
                   &    Predicted        &  Predicted \\
		           & 	Positive	     &  Negative\\
        \hline
        Actual     &  True Positive      &  False Negative \\
       Positive	   &  (TP)	             &  (FN) \\
        Actual     &  False Positive     &  True Negative  \\
        Negative   &  (FP)               &  (TN) \\
        \hline
	\end{tabular}
\end{table}

  Our goal is that the largest possible number of pulsar candidates are identified correctly while a minimal amount of mislabeled non-pulsar candidates are returned \citep{Morello2014}.
  We consider three metrics:
\begin{gather}
    Recall=\frac{TP}{TP+FN}\\
    Precision=\frac{TP}{TP+FP}\\
    F-Score =2\cdot \frac{Recall \cdot Precision}{Recall+ Precision}
\end{gather}

  Recall is the fraction of pulsars correctly classified,
  Precision means the proportion of pulsars properly classified in candidates which are classified as positive,
  and F-Score is the harmonic mean of precision and recall.
  For our goal, identifying the largest possible number of pulsars means that we want higher Recall,
  and returning a minimal amount of mislabeled noise indicates a high Precision.
  Recall and Precision are a pair of conflicting metrics.
  Normally, when Precision is high, Recall is usually low;
  when Recall is high, Precision is often low.
  We aim to get a better trade-off between Recall and Precision.
\section{Feature selection with relative importance}

\subsection{Feature relative importance}

  Based on tree models, we filter features further by feature relative importance.
  The ranking of relative importance represents the contribution of different features to the algorithm.

  Random Forest can output the feature importance by computing the mean decrease impurity \citep{Breiman1984} or mean decrease accuracy \citep{Breiman2001}.
  Mean decrease impurity is defined as the total decrease in node impurity (such as Gini impurity, information gain/entropy) averaged over all trees of the ensemble.
  These features with small decreases are less important
  Mean decrease accuracy measures the decrease in accuracy on OOB data when we randomly permute the values of each feature.
  The more important a variable is, the more its accuracy would decrease.
  Mean decrease accuracy provides similar variable ranking \citep{Verikas2011} but is computationally expensive \citep{Tuv2009}.
  So, we use mean decrease impurity(Gini impurity) to rank features' relative importance in the Random Forest method.
  For example, the initial Gini index of a node $t$  with data $D$ before splitting is
  \begin{equation}
    Gini(D)=1-p(c_+)^2-p(c_-)^2
  \end{equation}
  where $p(c_+)$ is the proportion of data points which have value $c_+$ for class variable, while $p(c_-)$ is the proportion which have value $c_-$ .
  Then if it is split by feature $A$,  the new Gini index will be
  \begin{equation}
    Gini_\_index(D,A)=\frac{D_1}{D}Gini(D_1)+\frac{D_2}{D}Gini(D_2)
  \end{equation}
  where $D_1$ and $D_2$ are data in right or left child node.
  The Gini decrease of node $t$  for feature $A$ is
  \begin{equation}
    I(t)=Gini(D)-Gini_\_index(D,A)
  \end{equation}
  Then total decrease for feature $A$ over all trees is
  \begin{equation}
    I_A=\frac{1}{M}\sum_{t=1}^{n}I(t)
  \end{equation}
  where $M$ is the number of trees in this Random Forest, $n$ means the number that used feature $A$ to split nodes. And $I_A$ measures the important of feature $A$.

  In XGBoost method, we chose another way to calculate feature scores after the boosted trees are constructed.
  They are based on the number of each feature that is used to split in each node. 
  Summing up the scores of each decision tree $N_t$, then averaged by the number of trees.
  \begin{equation}
    I_A=\frac{1}{M}\sum_{t=1}^{M}N_t
  \end{equation}
  The more an attribute is used to make key decisions with decision trees, the higher its score is \citep{Hastie2009}.

  Higher the feature scores are, more sufficient that feature helps in candidate classification.
  Such scores rely on data distribution of datasets, algorithms, and so on.
  As mentioned before, 30 features will be extracted from each candidate in every dataset.
  These features will get relative importance scores by Random Forest and XGBoost.

  Before analysing feature importance, we need to avoid the risk of overfitting.
  Overfitting is a problem in applied machine learning, which means that the model can not generalize well from training data to unseen testing data.
  It may fit the training dataset well while has poor generalization to other data.
  It occurs because models may have learned the noises from training data as useful features, which leads to poor model performance.
  To build up an estimate of how models might perform on unseen data and limit overfitting,
  we used the technique that holds back a sub-dataset in each dataset, which does not participate in selecting feature or tuning algorithms.

  To introduce in detail,  we randomly separated a testing sub-dataset at first from HTRU 1 or 2 dataset, respectively.
  They kept the same imbalanced ratios with initial sets.
  These data were independent and not involved in feature selection or model training, which were used only for testing models' performance in the next section.
  In the situation of  HTRU 1, test dataset contained 40\% candidates, while it had 20\% candidates in  that of HTRU 2. (The number of test data is consistent with that in \citet{Guo2017} or \citet{Mohamed2017}, for comparing results.)

\begin{figure}[t]
      \centering
      \includegraphics[width=\columnwidth]{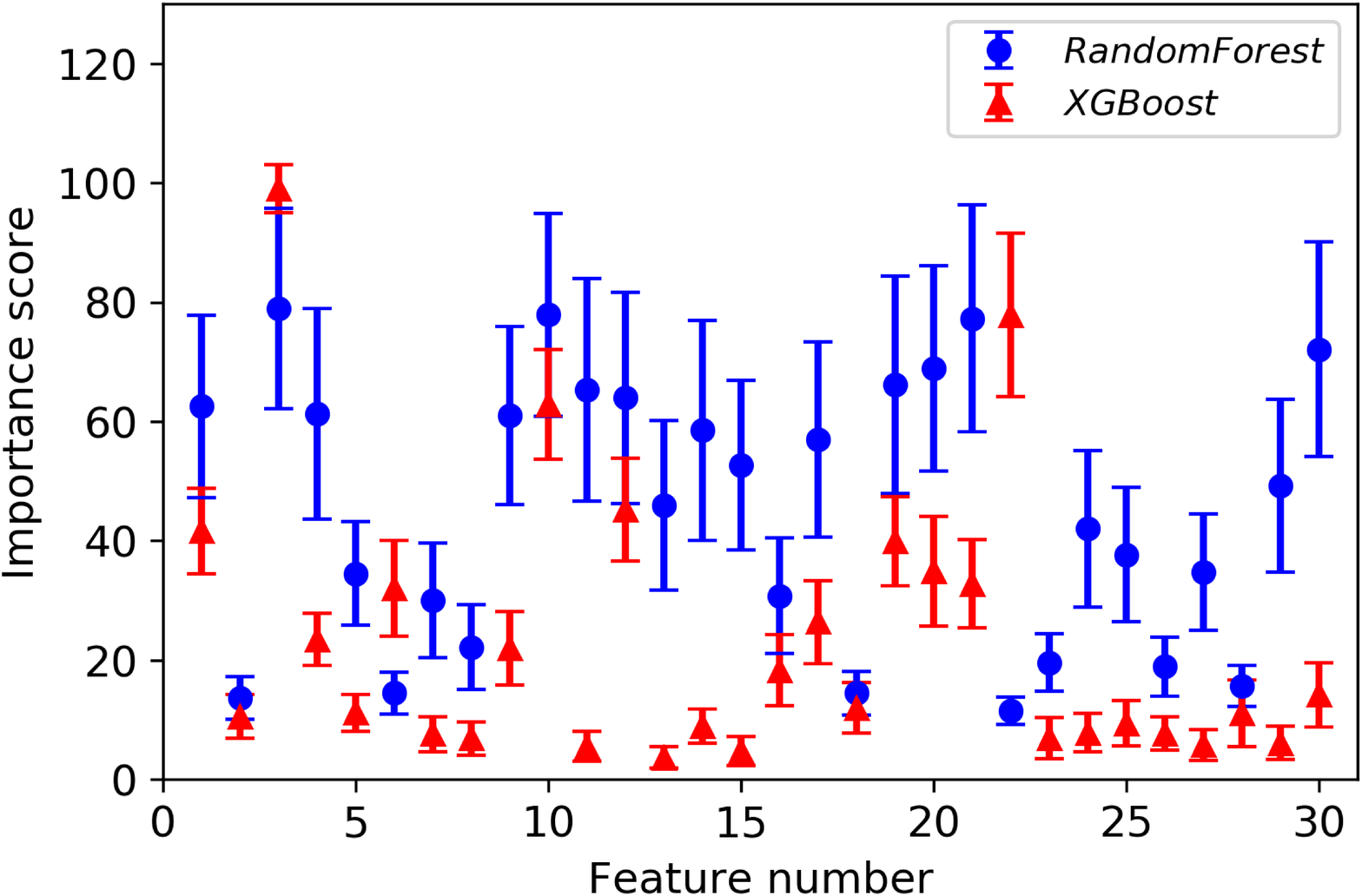}
      \caption{Feature scores of Random Forest and XGBoost averaged over 100 times in HTRU 1. }
      \label{fig:Figure 5}
\end{figure}

\begin{figure}[t]
      \centering
      \includegraphics[width=\columnwidth]{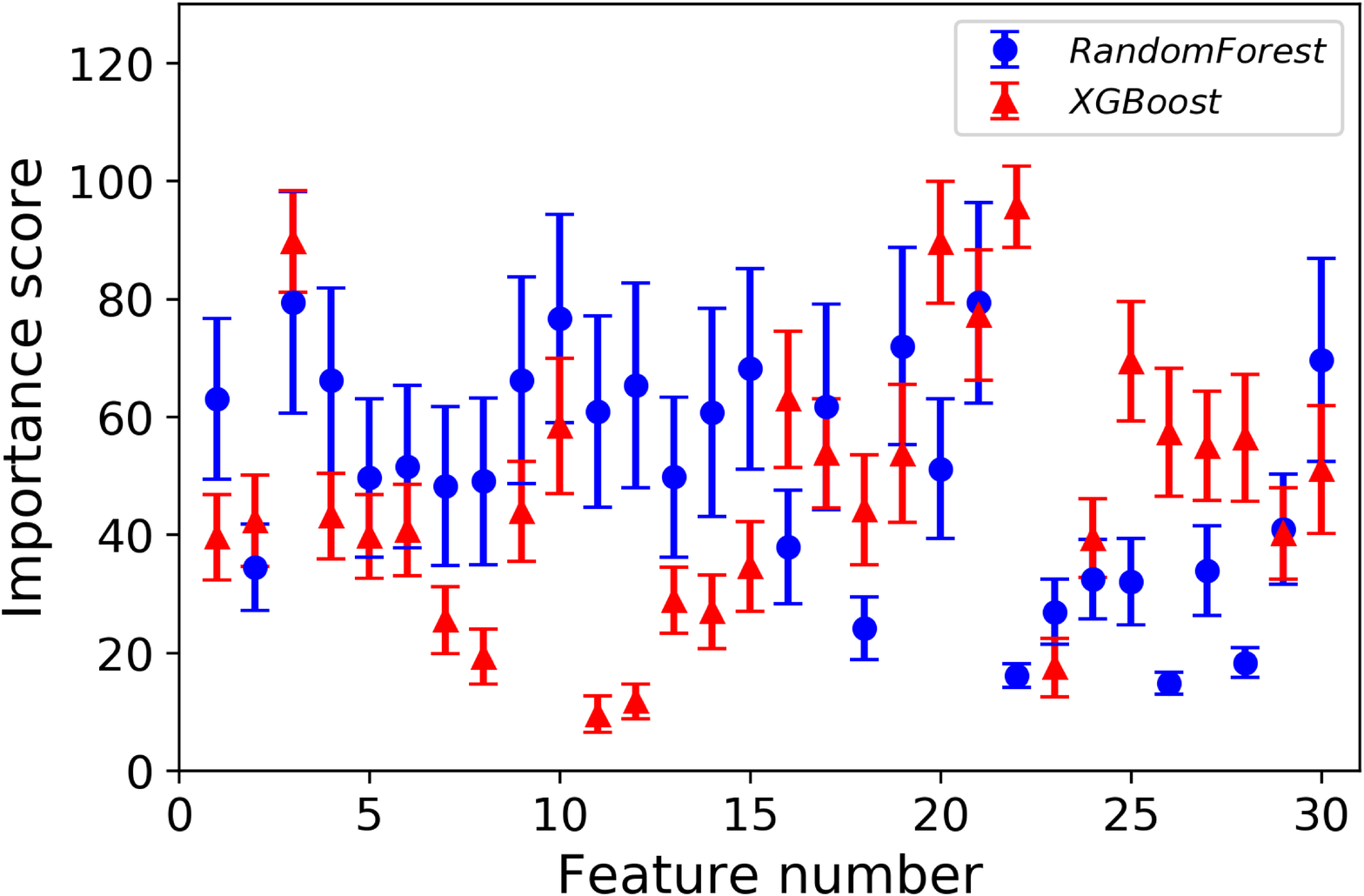}
      \caption{Feature scores of Random Forest and XGBoost averaged over 100 times in HTRU 2. }
      \label{fig:Figure 6}
\end{figure}

  On the other hand, the left sub-datasets(we called training sub-datasets) are used to train models and get feature scores.
  In this section, both two training sub-datasets were used to train Random Forest and XGBoost to get feature scores, respectively.
  In each training processing, sub-dataset was randomly divided into two equal parts: training and validation.
  With different training data, there may have fluctuation in scores. So we computed feature scores for 100 times.
  The scores from these tests are normalized and ranges between 0 to 100.
  The averaged normalized scores are shown in Table~\ref{tab:Table 3}.
  As comparisons, Figure~\ref{fig:Figure 5}  and Figure~\ref{fig:Figure 6}  show score differences in HTRU 1 and 2, respectively.


  We focused on the results from the test above, especially for the features with high scores ($\ge$ 70) or low scores ($\le$ 30).

  1) The fluctuation of higher score features are relatively huge (approximately $\pm$ 20)
  while that of lower score features tend to be close (approximately $\pm$ 10) to one value.

  2) For results from Random Forest model, their relative errors vary only a little:
  20\% to 34\% for HTRU 1 and 14\% to 30\% for HTRU 2.
  The situation is different when using XGBoost model:
  0.3\% to 63\% for HTRU 1 and 6\% to 29\% for HTRU 2.

  3)The number of higher score features are less than that of lower score features.
  For HTRU 1 dataset, Random Forest and XGBoost provided 4/9 and 1/21 high/low score features, respectively.
  For HTRU 2 dataset, these two algorithms provided 4/5 and 3/7 high/low score features, respectively.

  4) Feature 3 has high scores in two datasets with both Random Forest and XGBoost.
     Feature 10 and 21 have high scores in two datasets with Random Forest only.
     These three features should be more important in candidate classifications than others.

  5) 
     Feature 23 has low scores in two datasets with both Random Forest and XGBoost.
     Feature 18, 22, 26, and 28 have low scores in two datasets with Random Forest.
     Feature 7, 8, 11, 13, and 14 have low scores in two datasets with XGBoost.
     These ten features should be less important in candidate classifications than others.

\subsection{Feature selection with relative importance}

  After getting features relative importance, we selected features used to train algorithms in section 5.
  We extracted subsets of these 30 features with applying a Wrapper method \citep{Kohavi1997} in which selecting features directly optimize the values of Recall.
  What we did on Random Forest and XGBoost are as follows.

  1) Calculating the learner performance with present features.

  2) Removing the last one feature in present relative importance ranking.

  2) Then training the learner (Random Forest / XGBoost) with the left features subset, and comparing the new performance with that of previous features.

  3) If the Recall of new left features subset on testing data is improved (at least 0.001) than the Recall of the previous features,
     then repeating step 2 and 3; if not, stopping the feature selection. Those left features will be reserved and used in experiments for training and testing.

  In this section, every recall values were averaged by 30 times. In each time, training datasets were randomly separated into 40\%(training), 40\%(validating) and 20\%(testing), respectively. And the results with feature selection are shown in Table~\ref{tab:Table 4}.

\begin{table*}[ht]
	\centering
	\caption{Feature selection on HTRU 1 and  HTRU 2.}
	\label{tab:Table 4}
	\begin{tabular}{ccccc} 
        \hline
          Dataset              &  Method                & Features    &  Recall    &  Precision\\	
        \hline
         \multirow{8}*{HTRU 1} & \multirow{6}*{XGBoost} &  all        &  0.955      &  0.992 \\
                               &                        &without f13   &  0.958     &  0.994  \\
                               &                        &without f13 and f11    &  0.959     &  0.991  \\
                               &                        &without f13, f11 and f15     &  0.960     &  0.995  \\
                               &                        &without f13, f11, f15 and f29     &  0.961     &  0.993  \\
                               &                        & without f13, f11, f15, f29 and f27     &  0.958     &  0.994  \\
        \cline{2-5}

                               &\multirow{2}*{Random Forest} & all    &  0.966     &  0.979 \\
                               &                             & without f22     &  0.957     &  0.985  \\
        \hline
	    \multirow{5}*{HTRU 2}  & \multirow{3}*{XGBoost} &  all        &  0.864      &  0.956 \\
                               &                        & without f11     &  0.866     &  0.954  \\
                               &                        & without f11 and f12     &  0.864     &  0.955  \\
        \cline{2-5}
                               & \multirow{3}*{Random Forest} &  all     &  0.866      &  0.961 \\
                               &                              & without f26     &  0.869     &  0.960  \\
                               &                              & without f26 and f22     &  0.864     &  0.951  \\
        \hline
	\end{tabular}
\end{table*}

  In this way, for HTRU 1 training dataset, when we removed last one feature for Random Forest, the recalls are decreased. So we reserve all features.
  But removing feature 13, 11, 15 and 29 for XGBoost, made a better result.
  In HTRU 2 dataset, we removed feature 26 for Random Forest
  while removing feature 11 for XGBoost.
  The following experiments are based on these optimized features.

\section{Results}

  In experiments, we applied three algorithms,
  Random Forest, XGBoost, and a Hybrid Ensemble method which combined Random Forest and XGBoost with EasyEnsemble,
  to test training results and feature selection results.
  Our goal is identifying the largest possible fraction of pulsars while returning a minimal amount of mislabeled non-pulsar.
  So, we wish to obtain a high Recall and a high Precision at the same time.
  In Hybrid Ensemble method,
  we adjusted given threshold $\theta$,
  to make both Recall and Precision get relatively better trade-off values: the values of Recall are close to that of Precision.

\subsection{Evaluation with HTRU 1 dataset}

  For comparison,
  as noted above, test dataset contains 40\% HTRU 1 candidates, which is as same as in \citet{Guo2017}.
  When applying selected features (see Section 4) on the three algorithms,
  the training operations of each algorithm were repeated for 30 times, respectively.
  In each time, left sub-dataset were randomly divided into a group of two folds for training(30\%) and validation(30\%) to train the model.
  Then the trained model was used to classify test dataset.  
  Different divided partition may lead to differences in the model, so we averaged 30 times' testing results as the model's performance for analysing.
  The performance metrics on HTRU 1 dataset are shown in Table~\ref{tab:Table 5}.
  We also list results from \citet{Morello2014}, \citet{Lyon2016}, and \citet{Guo2017} in contrast.

  With cost-sensitive methods,
  both Random Forest and XGBoost can achieve good performance on Precision,
  which are 0.965 and 0.980, respectively.
  As for Recall, the performance of Random Forest is 0.965 (better than 0.955 in XGBoost) and still a little bit worse than that in \citet{Guo2017}, which is 0.966.

\begin{figure}[ht]
      \includegraphics[width=\columnwidth]{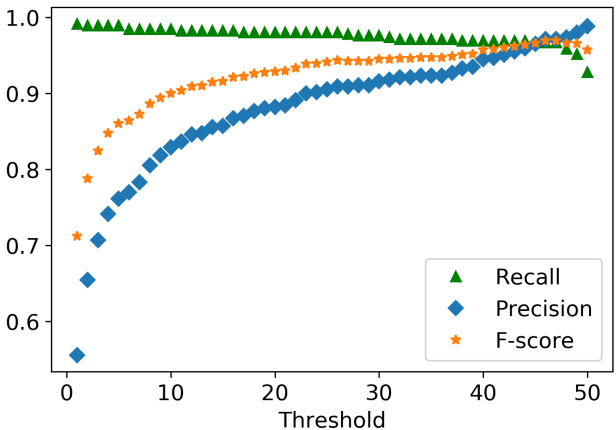}
      \caption{The Recall, Precision and F-Score performances of Hybrid Ensemble with changing voting threshold on HTRU 1 dataset.}
      \label{fig:Figure 7}
\end{figure}

\begin{figure}[ht]
      \includegraphics[width=\columnwidth]{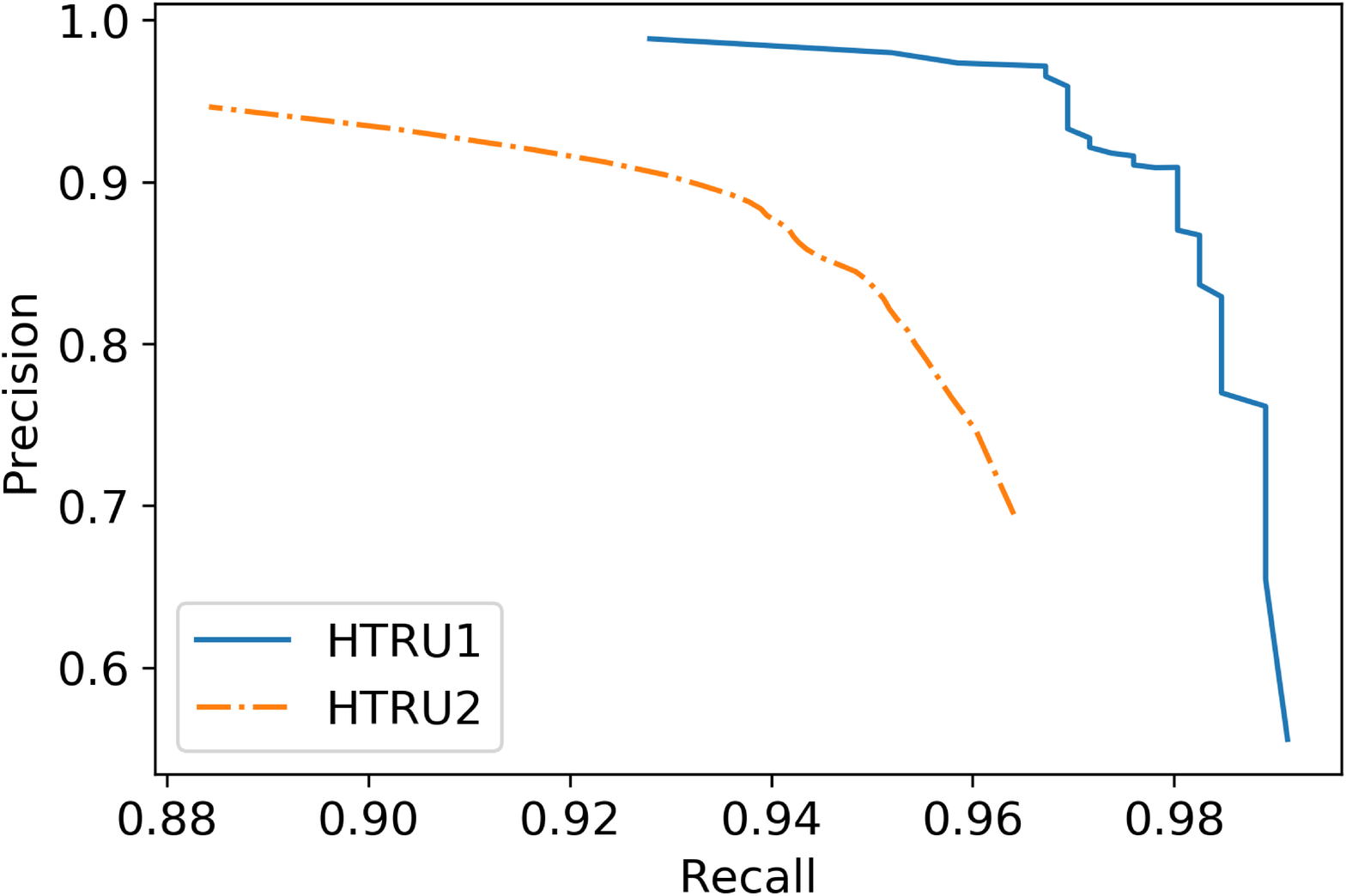}
      \caption{The Precision-Recall curve of Hybrid Ensemble on HTRU 1 and HTRU 2 dataset. }
      \label{fig:Figure 8}
\end{figure}

  For the Hybrid Ensemble,
  we extracted 50 subsets from majority class (training subset) randomly.
  Each subset only contains 5000 negative candidates (using the bootstrap method to make the sample distribution consistent).
  So, there are 50 corresponding base learners (25 for Random Forest and 25 for XGBoost, randomly) in EasyEnsemble.
  Then we use a voting algorithm to combine these 50 results.
  The performances with different voting threshold $\theta$ and Precision-Recall curve are shown in Figure~\ref{fig:Figure 7} and Figure~\ref{fig:Figure 8}.
  For the balance between Recall and Precision, we set the threshold $\theta$ to be 46.
  In this situation, Recall, Precision, and F-Score of Hybrid Ensemble are 0.967, 0.971, and 0.969.
  On Recall, Hybrid Ensemble improves 0.1\% over DCGAN-SVM method.
  On Precision, Hybrid Ensemble is worse than using XGBoost separately,
  but it still improves 0.6\% over DCGAN-SVM method.
  Hybrid Ensemble got a better trade-off between Recall and Precision on HTRU 1 dataset.

\subsection{Evaluation with HTRU 2 dataset}

  The testing subset got at last section contains 20\% HTRU 2 candidates, which was consistent with the operation in \citet{Mohamed2017}.
  The left candidates in HTRU 2 were used for both training(40\%) and validation(40\%).
  We also repeated the random division on the left subset to train each algorithm and test for 30 times, respectively.
  The performance metrics of classification on HTRU 2 dataset are averaged and showed in Table~\ref{tab:Table 6}.
  We also list the results of \citet{Lyon2016} and \citet{Mohamed2017} as a reference.

  With cost-sensitive methods, both Random Forest and XGBoost also achieved good performances on Precision (0.961 and 0.955, respectively).
  For Recall, the performance of XGBoost (0.873) is better than that of Random Forest (0.870), but both are worse than that of \citet{Mohamed2017} (0.942).

\begin{figure}[t]
      \includegraphics[width=\columnwidth]{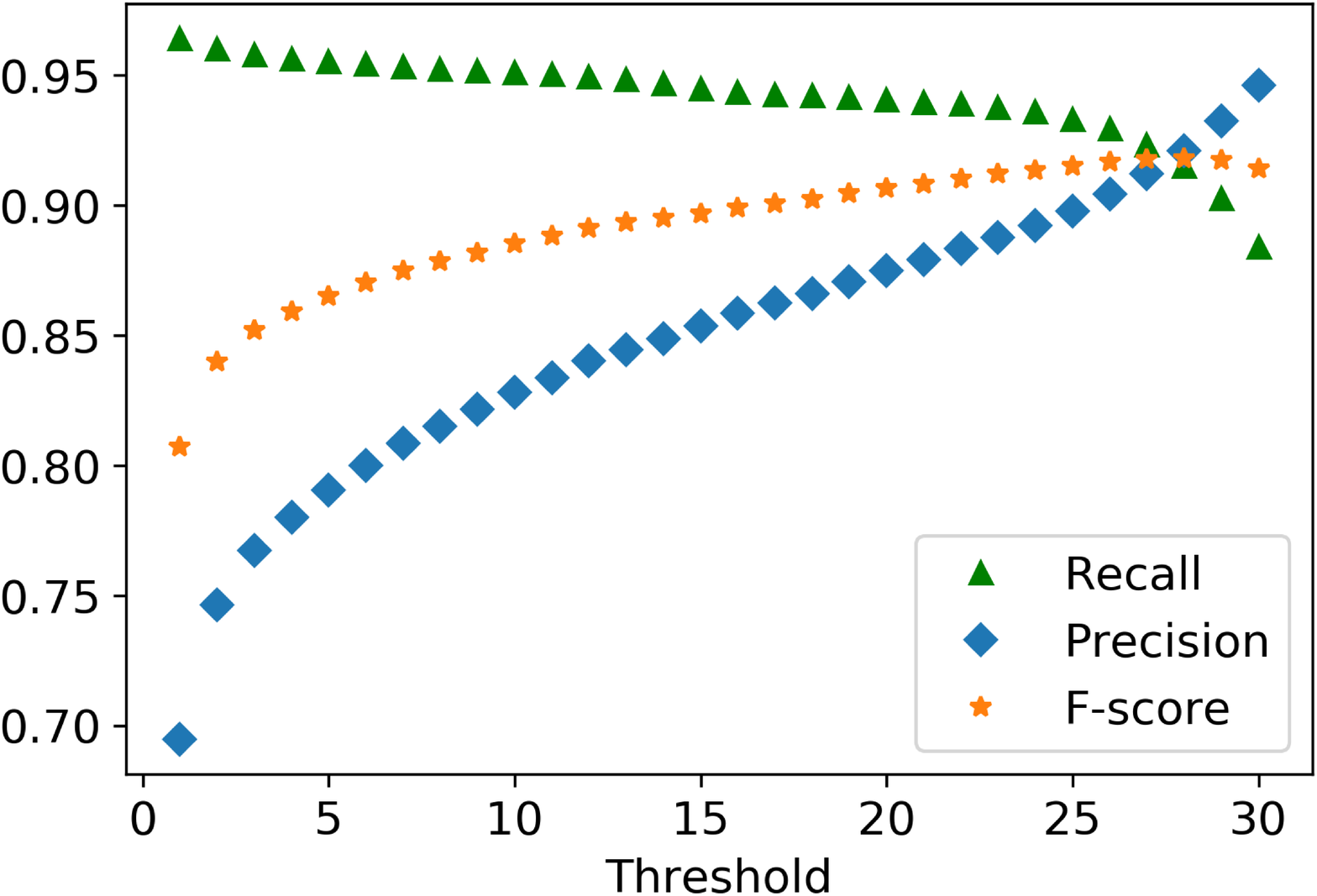}
      \caption{The Recall, Precision and F-Score performances of Hybrid Ensemble with changing voting threshold on HTRU 2 dataset.}
      \label{fig:Figure 9}
\end{figure}

  For the Hybrid Ensemble,
  there are 1300 negative candidates in each subset (30 in total) of majority class randomly.
  Correspondingly, we trained 30 base learners (15 for Random Forest and 15 for XGBoost) in EasyEnsemble.
  Then we used the same voting algorithm to combine these 30 results.
  Figure~\ref{fig:Figure 9} shows the performances with different thresholds.
  Figure~\ref{fig:Figure 8} shows its Precision-Recall curve.
  When threshold $\theta$ is $\theta=27$,
  both Recall and Precision can get a relatively better trade-off value.
  Recall, Precision, and F-Score of Hybrid Ensemble are 0.920, 0.917, and 0.918, respectively.
  Selecting $\theta=5$ as the voting threshold, we got metrics: 0.956, 0.813 and 0.879.
  Compared to Fuzzy-Knn method in \citet{Mohamed2017} (Precision is 0.808, very close to 0.813 in our work),
  the value of recall increased by 1.4\% while that of Precision is close.

\begin{table*}[t]
   \centering
   \caption{Performance of different methods on HTRU 1 dataset.}
   \label{tab:Table 5}
\begin{tabular}{cccc}
\tableline
   \textbf{Method}	& \textbf{Recall}	& \textbf{Precision}	& \textbf{F-Score} \\
\tableline

 \multirow{3}*{ANN\citep{Morello2014}} & 1	& 0.675	& 0.806 \\
                                              & 0.99 & 0.92 & 0.954  \\
                                              & 0.95 & 0.99	&0.970 \\
\hline
 GH-VFDT\citep{Lyon2016}	& 0.928	& 0.955	& 0.941 \\

\hline
 \multirow{2}*{DCGAN-SVM\citep{Guo2017}}  &0.963	&0.965	&0.964 \\
	                        &0.966	&0.961	&0.963\\
\hline	
  Random Forest	           &0.965	&0.965	&0.965  \\
  XGBoost	               &0.955	&\textbf{0.981}	&0.968 \\
  Hybrid Ensemble($\theta$=46)	&\textbf{0.967}	&0.971 &\textbf{0.969} \\
\hline

\end{tabular}
\end{table*}

\begin{table*}[t]
\centering
\caption{Performance of different methods on HTRU2 dataset.}
\label{tab:Table 6}
\begin{tabular}{cccc}
\hline
 \textbf{Method}	& \textbf{Recall}	& \textbf{Precision}	& \textbf{F-Score} \\
\hline
	GH-VFDT\citep{Lyon2016}	&0.829	&0.899	&0.862  \\
\hline
   Fuzzy-Knn\citep{Mohamed2017}	    &0.942	&0.808	&0.873 \\
\hline	
  Random Forest	&0.870	&\textbf{0.961}	&0.913 \\
  XGBoost	&0.873	&0.955	&0.912 \\
  Hybrid Ensemble($\theta$=27)	&0.920	&0.917	&\textbf{0.918} \\
  Hybrid Ensemble($\theta$=5)	&\textbf{0.956} 	&0.813	&0.879\\
\hline

\end{tabular}

\end{table*}


\section{Discussion}

\begin{figure}[t]
      \centering
      \includegraphics[width=\columnwidth]{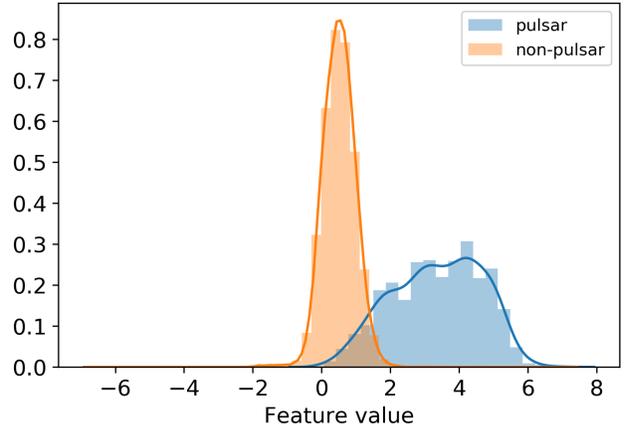}
      \caption{The distribution of skewness of the integrated pulse profile for pulsars vs. RFI on HTRU 1 dataset.}
      \label{fig:Figure 10}
  \end{figure}

  Importance score points the contributions of each feature to build the algorithm.
  The more an attribute is used to split the data,  the higher its relative importance.
  It is related to the performance measures (such as Gini purity, information gain) of each attribute split point.
  In a degree, it can describe the ability of an attribute to distinguish samples.
  Therefore, if the feature has better discrimination on this dataset, it will usually get a higher score.
  For example, the skewness of the integrated pulse profile is identified as one of the valuable features with a higher score.
  It should have a better ability to distinguish between pulsar and RFI.
  Figure~\ref{fig:Figure 10} shows the distribution of this attribute for pulsars vs. RFI on HTRU 1 dataset.
  We can find that it has relatively good linear separability for most data.
  Similarly, other useful features must have linear or nonlinear ability acting on training data, which makes them receive high scores.

  The importance scores are relative and rely on training data and feature sets.
  In our different training processes, the distribution of training pulsar candidates may be changed, due to the randomness of selecting candidates or features.
  The randomness leads to slight fluctuation in feature scores.
  Also, the presence of correlated features can cause fluctuations.
  As above mentioned, if the tree has used one of the correlated features, others will be rarely used and have a lower score.
  For other trees, the correlated features with low scores may play essential roles in splitting nodes and get a higher score.
  Therefore, fluctuations are reasonable.
  In a degree, the more important the feature is, the higher the fluctuation is.
  The averaged ranking tends to stable as the training repeated.

  These relative important features also have limitations.
  They may be valuable but not optimal.
  For example, S/N plays an important role in Random Forest or XGBoost, but it is known as a poor feature, which may be inclined to select strong pulsars while ignoring others.
  Therefore more feature optimization and design are needed.

  In this paper, for avoiding over-fitting, we randomly separated a test dataset at first from HTRU 1 and 2 dataset, respectively, before feature selection or model training.
  The testing results are only used to be tested in section 5. Due to the lack of labeled data, we did not do more cross-validation.
  In Random Forest and XGBoost, they have randomness, which increases the diversities and generalization performance.
  But, in the Hybrid Ensemble method, it is more complicated.
  Moreover, we tried to obtain Recall and Precision being traded off by changing thresholds, which may lead to over-fitting.
  In practice, we recommend using additional and independent data to optimize threshold in the Hybrid method.

\begin{figure*}[ht]
\subfigure[pulsar\_1038]{\label{fig:subfig:a}
\includegraphics[width=0.45\linewidth,height=7cm]{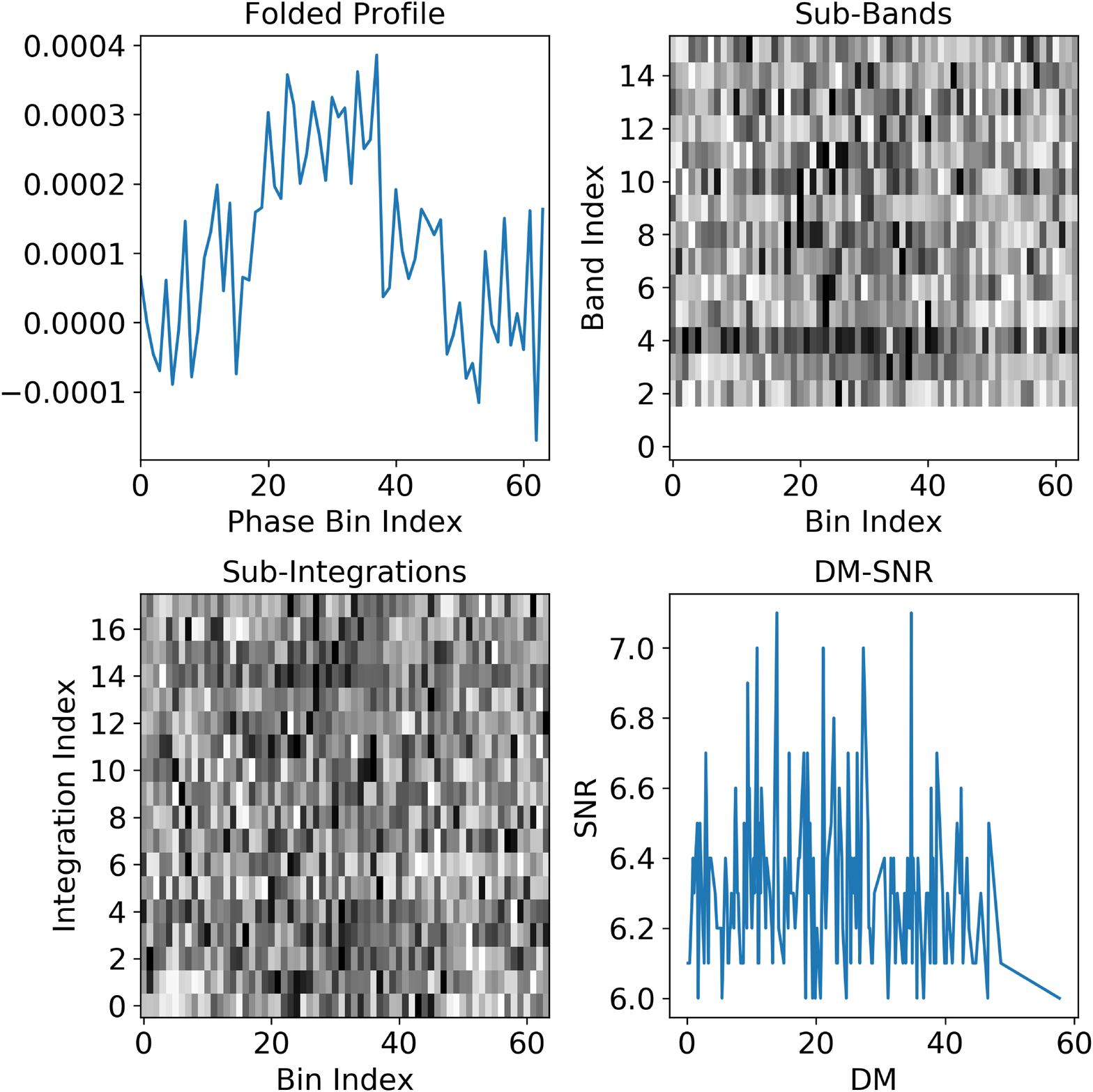}}
\hspace{0.01\linewidth}
\subfigure[pulsar\_1031]{\label{fig:subfig:b}
\includegraphics[width=0.45\linewidth,height=7cm]{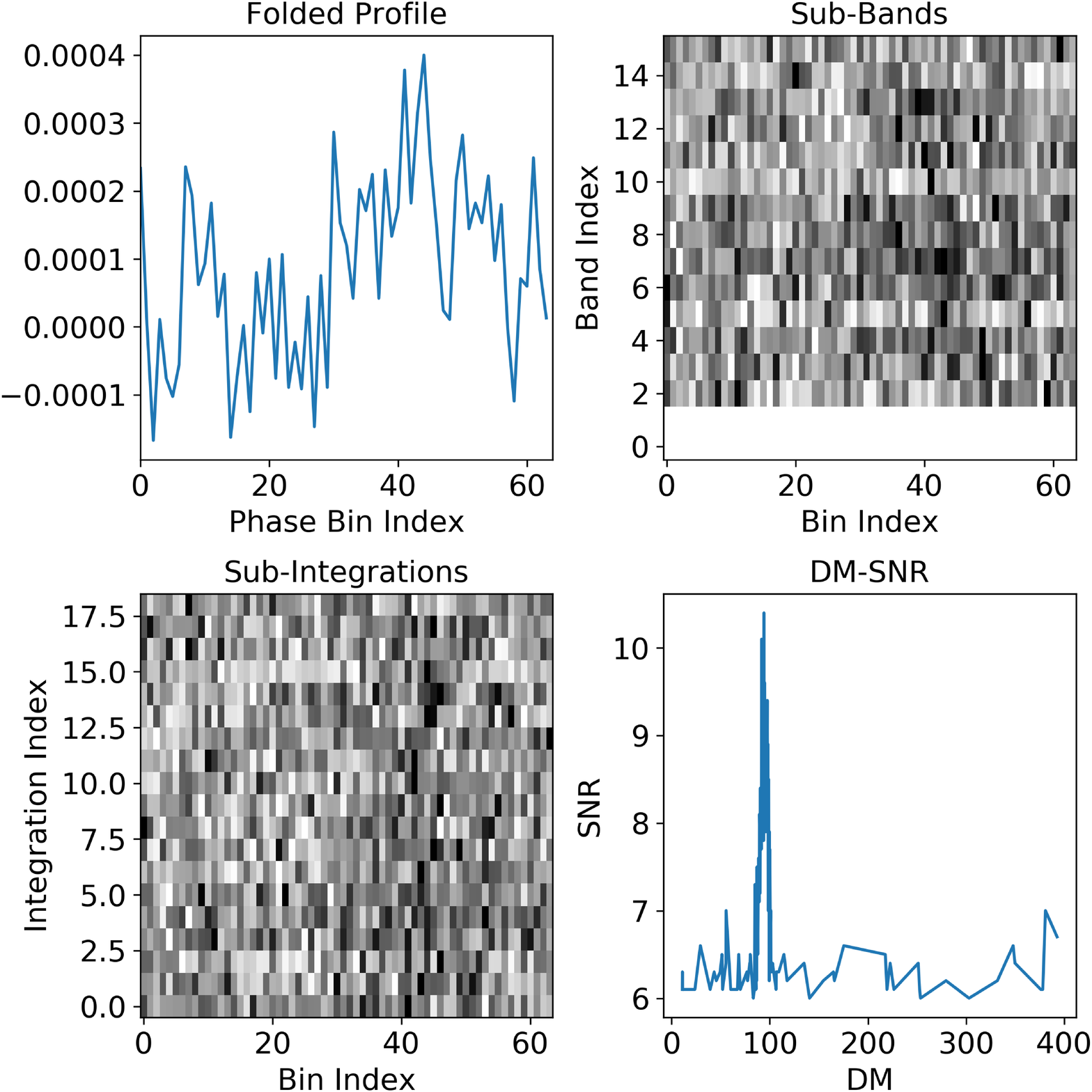}}
\vfill
\subfigure[pulsar\_0677]{\label{fig:subfig:c}
\includegraphics[width=0.45\linewidth,height=7cm]{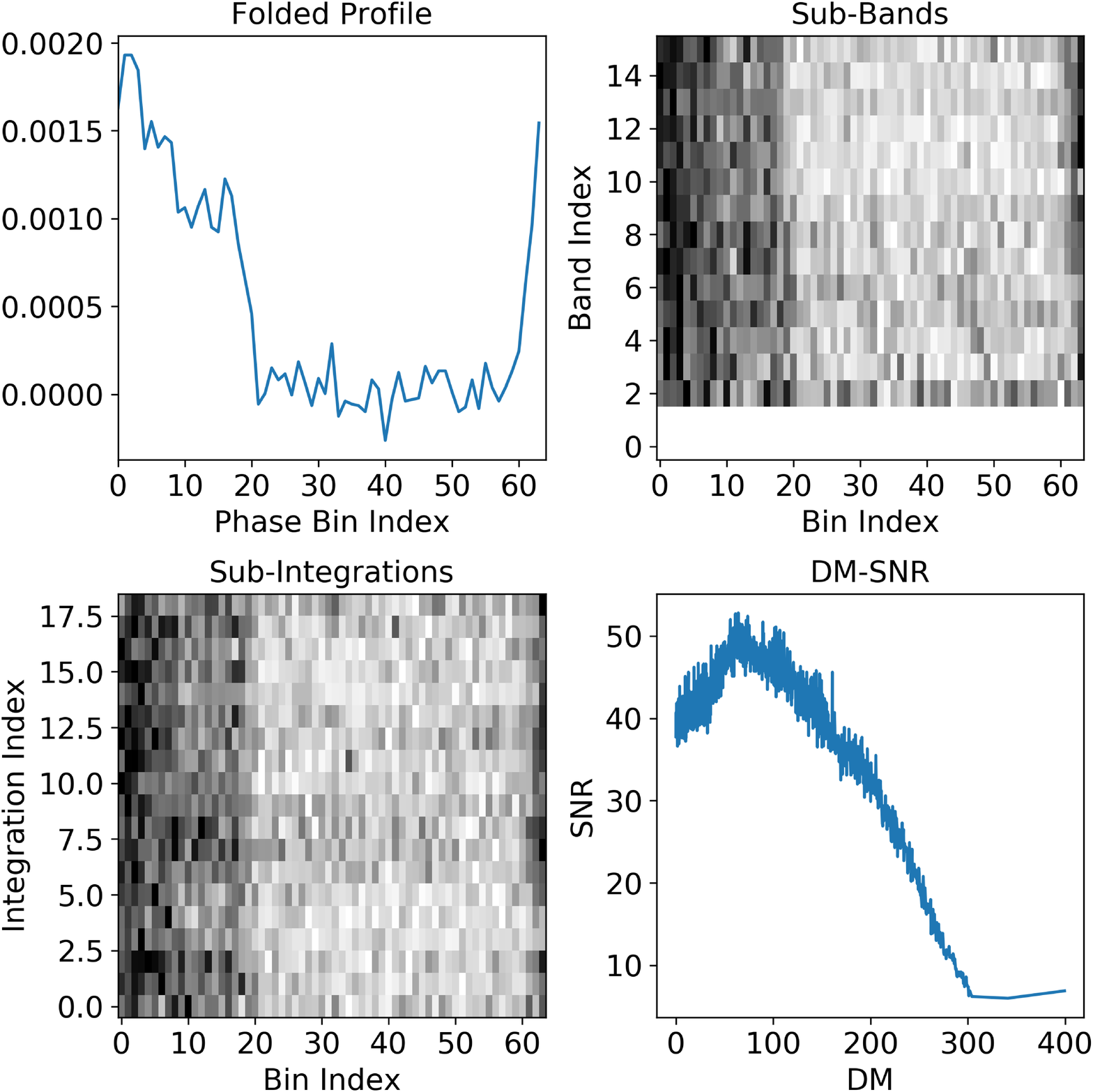}}
\hspace{0.01\linewidth}
\subfigure[pulsar\_0779]{\label{fig:subfig:d}
\includegraphics[width=0.45\linewidth,height=7cm]{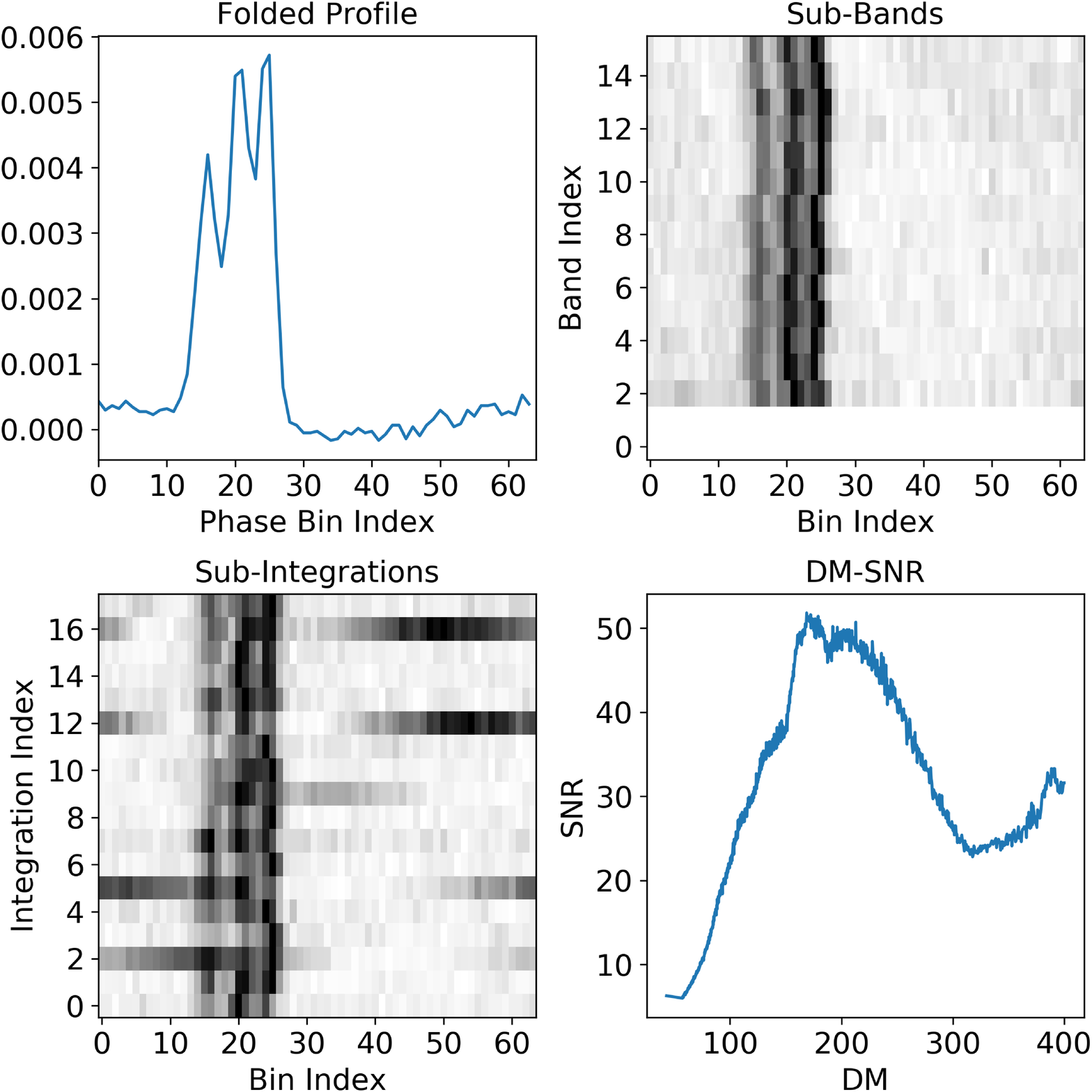}}
\vfill
\subfigure[pulsar\_0182 ]{\label{fig:subfig:e}
\includegraphics[width=0.45\linewidth,height=7cm]{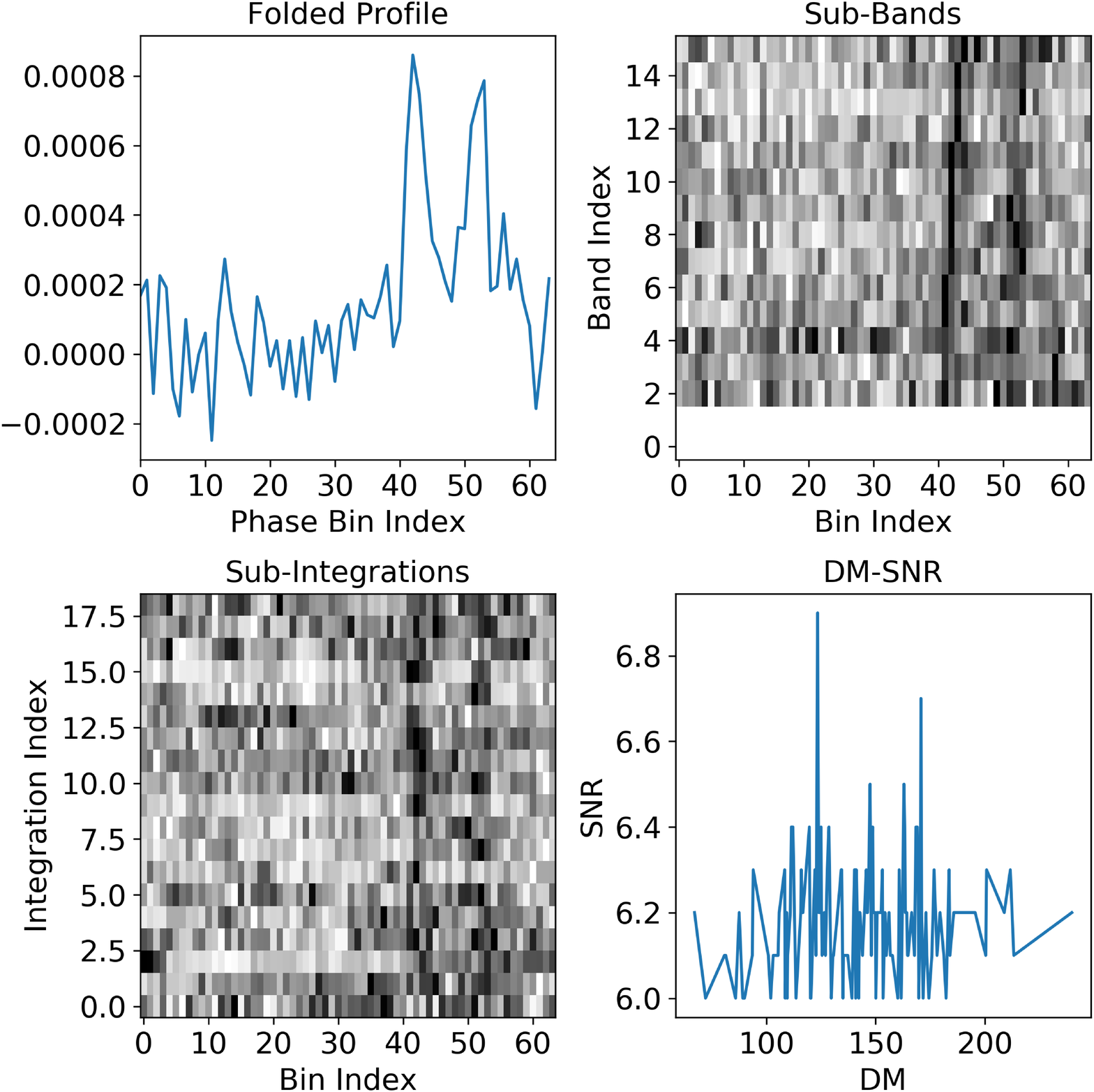}}
\hspace{0.01\linewidth}
\subfigure[pulsar\_0137]{\label{fig:subfig:f}
\includegraphics[width=0.45\linewidth,height=7cm]{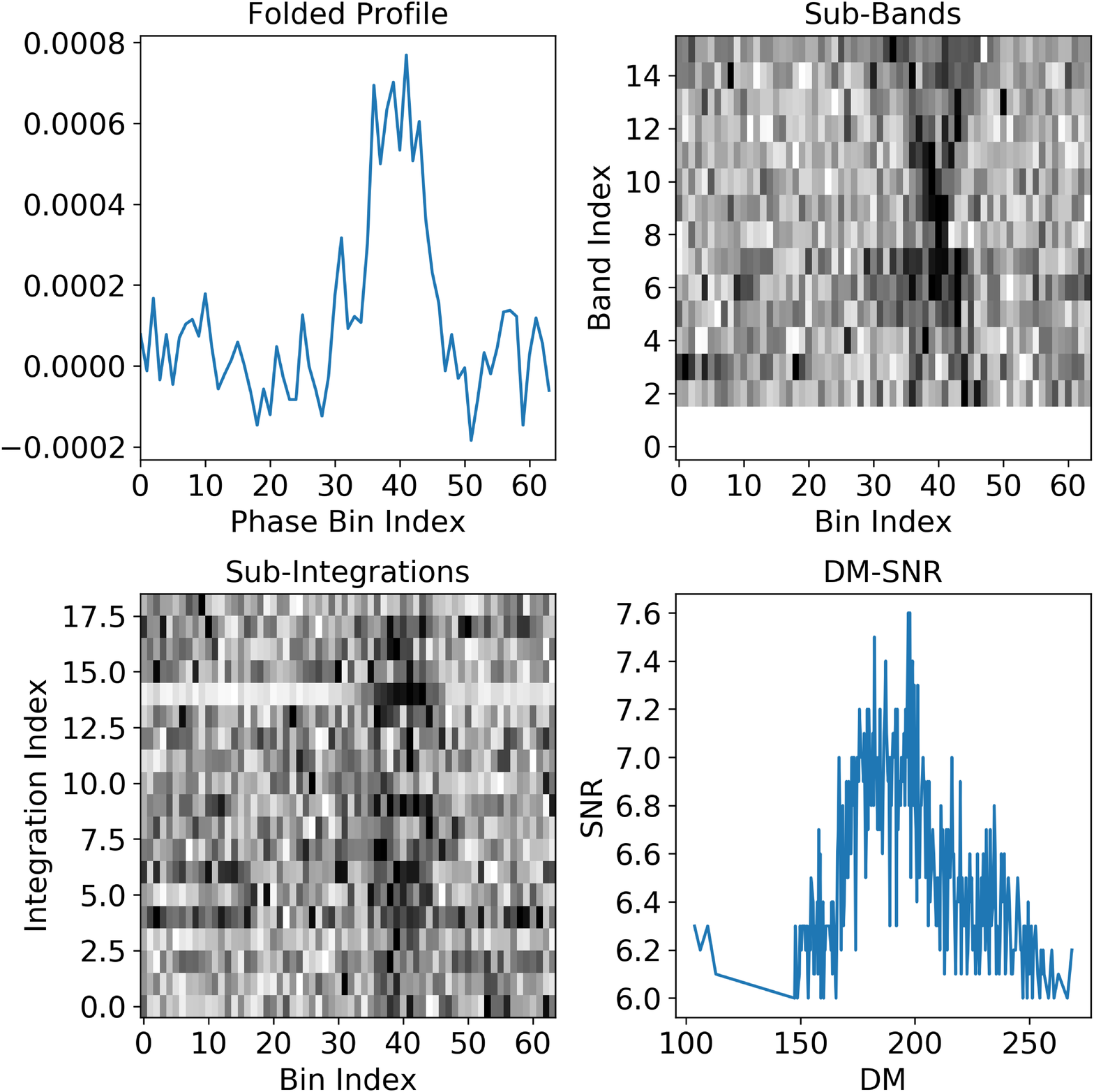}}
\caption{Some mislabeled candidates by Hybrid Ensemble in HTRU 1 dataset. (a)$\sim$ (f) are pulsar\_1038, pulsar\_1031, pulsar\_0677, pulsar\_0779 pulsar\_0182 and pulsar\_0137.}
\label{fig:Figure 11}
\end{figure*}

  The Hybrid Ensemble method performs better on two datasets.
  There are still several problems that need further improvements.
  Figure~\ref{fig:Figure 11} shows some typical mislabeled candidates in HTRU 1 dataset.
  We can find that pulsar candidates with a relative wide pulse or multi-peaks are more likely to be mislabeled.
  One possible reason is that these mislabeled pulsar candidates differ from normal pulsars.
  For example, they have wide pulse widths making them more like to be RFIs.
  Besides, some features that we used may be sub-optimal for pulsars with wide integrated pulse profiles, which may have an adverse effect on the ML systems' performance\citep{Tan2018}.
  The other reason is that most of the features we used are extracted from the folded profile plot and the DM-SNR curve.
  When pulsars are interfered by some noises, the pulses in the folded profile may become unobvious, this makes pulsars mislabeled.
  If we design some new features to characterize sub-integrations plots or sub-bands plots well,
  it may improve the performance and make the algorithm more practical in application.
  In \citet{Tan2018}, authors introduced 12 new features, which improved performance on candidates with relative wide pulses.
  We will learn from their work and try to make some progress in our future work.

\section{Conclusion}
  In this work, we combined the features of \citet{Lyon2016} with \citet{Bates2012},
  and used feature relative importance ranking to select them.
  Then based on selected features, we applied three algorithms: Random Forest, XGBoost, and a Hybrid Ensemble method in two imbalanced datasets of HTRU.
  The conclusions are shown as follows:

  1) The skewness of the integrated pulse profile is one of the most important features in XGBoost.
     In our Random Forest model, the skewness of the integrated pulse profile, Chi-Squared value for sine-squared fit to amended profile, and Best S/N value play important roles.

  2) In the Hybrid Ensemble, we combined Random Forest and XGBoost with EasyEnsemble.
     It is effective for this imbalance problem and makes a better trade-off between Recall and Precision in experiments.

  3) In test data of HTRU 1 dataset, Recall, Precision, and F-Score of the Hybrid Ensemble method are $0.967$, $0.971$, and $0.969$, respectively.
     In test data of HTRU 2 dataset,  these values are $0.920$, $0.917$, and $0.918$, respectively.
     Both of the results are better than previous methods used in \citet{Guo2017} and \citet{Mohamed2017}.

\section*{Acknowledgements}
%

  We would like to thank Vincent Morello and Robert Lyon for providing datasets and feature extraction program scripts publicly available,
  which are very helpful to our research.



\bibliography{AAS-bibtex}
\bibliographystyle{spr-mp-nameyear-cnd}

\end{sloppypar}
\end{document}